# Substrate-supported thermometry platform for nanomaterials like graphene, nanotubes, and nanowires

Zuanyi Li,[1,2] Myung-Ho Bae,[3] and Eric Pop[1*]

[1]*Electrical Engineering, Stanford University, Stanford, California 94305, USA*
[2]*Department of Physics, University of Illinois, Urbana-Champaign, Illinois 61801, USA*
[3]*Korea Research Institute of Standards and Science, Daejeon, 305-340, South Korea*

We demonstrate a substrate-supported thermometry platform to measure thermal conduction in nanomaterials like graphene, with no need to suspend them. We use three-dimensional simulations and careful uncertainty analysis to optimize the platform geometry and to obtain the sample thermal conductivity. The lowest thermal sheet conductance that can be sensed with <50% error is ~25 nWK$^{-1}$ at room temperature, indicating applicability of this platform to graphene or polymer thin films, nanotube or nanowire arrays, even a single Si nanowire. The platform can also be extended to plastic substrates, and could find wide applicability in circumstances where fabrication challenges and low yield associated with suspended platforms must be avoided.

[*]Contact: epop@stanford.edu

Understanding heat flow in nanomaterials is important both for fundamental knowledge and practical applications.[1-4] Different measurement techniques have been developed to probe nanoscale thermal conduction, such as the 3ω method, scanning thermal microscopy, time-domain thermoreflectance, and various bridge platforms.[5] Among them, using a microfabricated suspended bridge can directly measure in-plane heat flow through nanostructures by electrical resistance thermometry.[6-9] This suspended thermometry has good accuracy and has been widely used for nanofilms,[9-11] one-dimensional (1D) materials like nanowires (NWs)[7, 8] and nanotubes,[12, 13] as well as two-dimensional (2D) materials like hexagonal boron nitride[14] and graphene.[15, 16] A major drawback of this platform is that the fabrication is complicated and the test sample has to be either fully suspended[6-15] or supported by a suspended dielectric (e.g. $SiN_x$) membrane.[16, 17] This limits the diversity of measureable materials and makes the suspended platform fragile.

To overcome these limitations, substrate-supported platforms could be preferred and have been recently employed in thermal studies of Al nanowires,[18] encased graphene,[19] and graphene nanoribbons in our previous work.[20] Nanomaterials are almost always substrate-supported in nanoscale electronics;[21] thus, a substrate-supported platform has the advantage of testing devices including extrinsic substrate effects,[4, 22] which could be different from their intrinsic thermal properties (probed by suspended platforms).[23-25] Substrate-supported thermometry platforms could also readily be incorporated in industrial mask designs and fabrication processes as thermal test structures in addition to existing electrical test structures. Thus, measuring the thermal conduction of nanomaterials on a substrate is crucial from a practical viewpoint.

In this work, we critically examine the applicability and limitations of nanoscale thermal measurements based on a substrate-supported platform utilizing electrical resistance thermometry. As a prototype, the thermal conductivity of graphene on a $SiO_2$/Si substrate was experimentally tested. The fabrication of this supported platform is much easier than that of suspended platforms, but as a trade-off, the thermal conductivity extraction is slightly more challenging and must employ a three-dimensional (3D) heat flow simulation of the test structure. Through careful uncertainty analysis, we find that the supported platform can be optimized to improve the measurement accuracy. The smallest thermal sheet conductance that can be measured by this method within a 50% error is ~25 nWK$^{-1}$ at room temperature, which means the supported platform can be applied to nanomaterials like carbon nanotube (CNT) networks, CNT or NW arrays, and even a single Si NW. Additionally, it is suitable for materials which cannot be easily suspended, like many polymers, and the substrate is not limited to $SiO_2$/Si but can be extended to other substrates such as flexible plastics.

Figure 1(a) shows a scanning electron microscopy (SEM) image of a typical supported thermometry platform, here applied to a monolayer graphene sample. (In general, the sample to be measured is prepared on a $SiO_2$/Si substrate, though this is not always necessary, as we will show below.) Then, two parallel, long metal lines with at least four probe arms are patterned by electron-beam (e-beam) lithography as heater and sensor thermometers. If the sample is conductive (here, graphene), then the heater and sensor must be electrically insulated by a thin $SiO_2$ layer, as seen in the Fig. 1(b) cross-section. To perform measurements, a DC current is passed through one metal line (heater) to set up a temperature gradient across the sample, and the electrical resistance changes of both metal lines (heater and sensor) due to the heating are monitored. After temperature calibration of both metal line resistances the measured *changes* in resistance ($\Delta R$) can be converted into changes in temperature of the heater and sensor, $\Delta T_H$ and $\Delta T_S$, as a function of heater power $P_H$.

As the sample is not suspended, a control experiment should be performed after removing the exposed parts of the sample and repeating the above measurements to independently find the thermal properties of the heat flow path through the contacts and substrate. Etching away the sample is important, rather than simply performing the measurement without the original sample, because it preserves the sample portion beneath the heater and sensor electrodes, i.e., the same contact resistance in both configurations. Using this approach, in a previous work,[20] we obtained the thermal conductivity of the underlying $SiO_2$ within 5% error of widely known values, which also helps support the validity of this approach.

We note that the fabrication of the supported platform can be performed with greater yield than that of suspended platforms, but the thermal conductance between the heater and sensor cannot be obtained analytically as in the suspended case, due to non-negligible heat leakage into the substrate. Therefore, numerical modeling of such heat conduction must be employed to extract the thermal conductivity of the sample. For comparison, we considered both 2D and 3D finite element models of the sample, which are implemented by a commercial software package (COMSOL). In the 2D model, only the cross-section of the platform is simulated, and the Si substrate size is chosen as $2L_S \times L_S$ [Fig. 1(b)]. In the 3D model, half of the platform needs to be simulated due to the symmetry plane, which bisects the region of interest, and the Si substrate size is chosen as $2L_S \times L_S \times L_S$ [Fig. 1(c)]. To perform the simulation, in both 2D and 3D models, the bottom and side boundaries (except symmetry plane in 3D) of the Si substrate are held at the ambient temperature, *i.e.*, isothermal boundary condition. Other outer boundaries of the whole structure are treated as insulated, *i.e.*, adiabatic boundary condition. Joule heating is simulated by applying a power density within the heater metal, and the calculation is performed to obtain the temperature distribution in steady state, as shown in Figs. 1(b) and 1(d) for the 2D and 3D models, respectively. After calculating the average temperature rises in the measured segments of the heater and sensor, we obtain the simulated values of $\Delta T_H$ and $\Delta T_S$ vs. $P_H$. Then, we can match these with the measured values by fitting the thermal conductance $G$ of the test sample between the heater and sensor. The thermal conductivity of the sample, $k = GL/A$, can then be extracted. Here, $L$ is the sample length, *i.e.*, the heater-sensor separation and $A$ is the cross-sectional area, *i.e.*, the sample width $W$ times thickness $h$ [see Figs. 1(b) and 1(d)].

To correctly obtain the sample $k$, the simulated size $L_S$ of the Si substrate must be carefully chosen because the real Si chip is ~0.5 mm thick and several mm wide. The chip dimensions are semi-infinite compared to the small heating region (~10 μm) and cannot be fully included due to computational grid limits. Thus, the simulated $L_S$ should be large enough to model the heat spreading and yield a converged value of the extracted $k$. For the 2D and 3D models, the extracted $k$ as a function of simulated $L_S$ from our graphene measurement[20] at 270 K are shown in Fig. 2(a). Here, for the 3D model, we further considered two cases: heater and sensor with probe arms [as shown in Fig. 1(d)] and without probe arms (see Fig. S1 in the supplementary material[26]), because the latter has a more direct correspondence to the 2D model (reflected by the similar extracted $k$ for small $L_S$). Comparing these two cases also allows us to test how many details of the electrode geometry should be included.

It is clear that the $k$ extracted by the 2D model continues decreasing as the simulated $L_S$ increases, whereas the two 3D models give converged $k$ when the simulated $L_S$ is sufficiently large (≥50 μm). The 2D model is insufficient because it neglects the heat spreading along the *y*-direction perpendicular to the 2D plane. Although the heater and sensor length (~10 μm or similar to the sample width) are long compared to their separation $L_{HS}$ (~0.5 μm), we find that 3D heat spreading about 10 μm away from the heating center cannot be neglected [Fig. S1(b) in the supplementary material[26]]. As the simulated $L_S$ increases, the neglected heat spreading in the y-direction becomes stronger in the 2D model. Thus, we find that the 3D simulation is preferable in order to fully capture all heat spreading effects due to the finite size of the sample.

Figure 2(a) also shows that simulating the effect of heat loss through the voltage probe arms is necessary. As shown in Fig. 2(b), if the same sample $k$ is used, the simulated temperature rise along the sensor for the "3D with probe" case (solid red line) is lower than that for the "3D without probe" case (solid black line), and in the former case there are temperature dips at the points where the probe arms are connected due to heat leakage through them. Figure 2(c) shows the extracted $k$ increases and saturates gradually as the simulated length of the probe arms $L_{probe}$ increases, indicating $L_{probe} \geq 1.5$ μm is sufficiently long to catch its effect and this converged value provides a correct $k$ of the sample.

Next, we turn to the estimation of uncertainty in the extracted thermal conductivity $k$, which can be accomplished by the classical partial derivative method: $u_k/k = [\Sigma_i (s_i \times u_{xi}/x_i)^2]^{1/2}$, where $u_k$ is the total uncertainty of extracted thermal conductivity $k$, $u_{xi}$ is the estimated uncertainty for each input parameter $x_i$ of the simulation, and the sensitivity $s_i$ is defined by $s_i = (x_i/k)\partial k/\partial x_i = \partial(\ln k)/\partial(\ln x_i)$. The sensitivity is evaluated numerically by giving a small perturbation for each input parameter around its typical value and redoing the extraction simulation to obtain the new $k$.[19] To highlight the relative importance of each input parameter, we define its absolute contribution as $c_i = |s_i| \times (u_{xi}/x_i)$, and relative contribution as $c_i^2/\Sigma c_i^2$. As an example, the calculated sensitivities and uncertainty analysis for the extracted $k$ in Fig. 2(c) are shown in Table S1 of the supplementary material.[26] The total uncertainty in this case is ~21%, and it mainly arises from the contributions ($c_i > 5\%$) of the thermal conductivity of bottom $SiO_2$ ($k_{ox}$), thermal boundary resistance (TBR) of the $SiO_2$/Si interface ($R_{oxs}$), measured sensor response ($\Delta T_S/P_H$), thermal conductivity of Si substrate ($k_{Si}$), and heater-sensor midpoint separation ($L_{HS}$). TBRs of the sample/$SiO_2$ interface ($R_{gox}$) and top $SiO_2$/metal interface ($R_{mox}$) are included in the uncertainty analysis, but their contributions are small ($c_i \leq 2\%$)[26] and are not shown in the subsequent discussion.

The accuracy of our supported thermometry platform can be optimized through two important geometric parameters: (i) the center-to-center distance between the heater and sensor ($L_{HS}$) and (ii) the bottom insulator (here oxide) thickness ($t_{box}$) [see Fig. 1(b)]. If $L_{HS}$ is too large, then too much of the heater power is dissipated into the substrate; if it is too short, then the temperature drop between heater and sensor is not large compared with the temperature variation under the heater/sensor. If $t_{box}$ is too thin, significant heat leakage will occur into the substrate; if it is too thick, then its lateral thermal conductance will dominate the heat flow between heater and sensor, overwhelming that of the supported sample.

The optimized values of $L_{HS}$ and $t_{box}$ can be found by monitoring the uncertainty change (due to sensitivity change) of extracted sample $k$, and the results are shown in Figs. 3(a) and 3(b). Here we consider $T = 300$ K and thin film sample of thermal sheet conductance $G_\square = kh = 100$ nWK$^{-1}$, and assume the sample is partly etched off so that only the part between the heater and sensor is preserved [see the inset of Fig. 3(b)].[27] The optimization is calculated at heater and sensor linewidth $D_{met} = 200$ nm, and only the input parameters whose contributions $c_i$ are larger than 2% are included. From the estimated total uncertainty, we find that the optimized values of $L_{HS}$ and $t_{box}$ are ~600 nm and ~300 nm, respectively, leading to the minimized uncertainty ~18%. By looking at the uncertainty contributed by each input parameter, it is clear that the optimization is achieved mainly due to the competition between the measured sensor response ($\Delta T_S$) and the role of the bottom oxide ($k_{ox}$), as we explained above. Additional calculations (not shown) indicate that narrower heater and sensor linewidths (~100 nm) give almost the same optimized values of $L_{HS}$ and $t_{box}$, but slightly lower total uncertainty.

After optimizing our supported thermometry platform, the next question concerns the smallest in-plane thermal conductance that can be sensed by this method. To address this, we calculate the uncertainty change of extracted thermal conductivity as a function of the sample thermal sheet conductance $G_\square = kh$. When the sample is partly etched to match the dimensions of the heater-sensor width and spacing [inset of Fig. 3(b)] the results at $T = 300$ K are shown in Fig. 3(c). As expected, the estimated measurement uncertainty increases as $G_\square$ decreases. If we set the maximum uncertainty to ~50%, the sensible range of $G_\square$ enabled by this platform is $G_\square > 25$ nWK$^{-1}$ [the blue region in Fig. 3(c)]. For samples that are not etched to conform to the heater-sensor width and separation [e.g., Fig. 1(a)], the uncertainty is slightly higher, and the smallest sensible $G_\square$ within a 50% error is ~32 nWK$^{-1}$ [see Fig. S2(c) in the supplementary material[26]]. This requirement could be satisfied in most thin film materials, such as polymer films ($k > 0.2$ Wm$^{-1}$K$^{-1}$ and $h > 200$ nm)[28] and CNT networks ($k > 20$ Wm$^{-1}$K$^{-1}$ and $h > 2$ nm).[29, 30]

We emphasize that this supported thermometry platform can be applied not only to thin films but also to *arrays* of quasi-one-dimensional materials [inset of Fig. 3(c)]. In our previous work,[20] we had shown its application to graphene nanoribbon arrays. Here, we give estimation of minimum array density required to apply the platform to carbon nanotube and Si nanowire arrays. For single-wall CNT arrays, assuming array density $p$ (the number of CNTs per unit width), the equivalent thermal sheet conductance is $G_\square = k(\pi d\delta)p$, where $k$, $d$, and $\delta$ are the thermal conductivity, diameter, and wall thickness of single-wall CNTs, respectively. Then, the array density is given by $p = G_\square/(k\pi d\delta)$. Considering single-wall CNTs with $k = 1000$ Wm$^{-1}$K$^{-1}$, $d = 2$ nm, and $\delta = 0.34$ nm, as well as $G_\square = 25$ nWK$^{-1}$ (the best case), we obtain the density required for CNT array measurements is $p \geq 12$ μm$^{-1}$. This CNT array density is achievable experimentally today, as some studies[31, 32] have demonstrated CNT densities up to ~50 μm$^{-1}$.

For NW arrays and for thicker multi-wall CNTs, the array density is given by $p = G_\square/(k\pi d^2/4)$, where $k$ and $d$ are the thermal conductivity and diameter of the NWs, respectively. For 20 nm diameter Si NWs[33] with $k \approx 7$ Wm$^{-1}$K$^{-1}$, the required array density is $p \geq 11$ μm$^{-1}$; for 50 nm diameter Si NWs, smooth and rough edges lead to $k \approx 25$ and $2$ Wm$^{-1}$K$^{-1}$, respectively,[33,34] and the required array density is $p \geq 0.5$ μm$^{-1}$ and $6$ μm$^{-1}$. Nanowire arrays can be fabricated much denser than these required densities.[17, 35] The above estimations indicate that the supported thermometry platform can be easily applied to CNT arrays and Si NW arrays with both smooth and rough edges. In addition, such arrays do not require uniform spacing.

We note that the minimum array density for 50 nm diameter smooth Si NWs is very low (~0.5 μm$^{-1}$), which implies that it is possible to measure a single Si NW by using this platform. To confirm this idea, we performed the simulation with just one NW between the heater and sensor [inset of Fig. 4(a)]. To achieve the best measurement accuracy, we first optimize the dimensions of the heater and sensor, that is, the midpoint distance between them ($L_{HS}$) and the distance between two voltage probe arms ($D_{pV}$) [inset of Fig. 4(a)]. The calculated uncertainty contributed from the measured sensor temperature rise ($\Delta T_S$) as a function of $L_{HS}$ and $D_{pV}$ is shown in Fig. 4(a). In the calculation, the bottom oxide thickness ($t_{box}$) and electrode linewidth [$D_{met}$, see Fig. 1(b)] are chosen as 300 nm and 200 nm, respectively, and a Si NW with $d = 50$ nm and $k = 25$ Wm$^{-1}$K$^{-1}$ is used. The minimum of the uncertainty indicates the optimized structure is $L_{HS} = 600$ nm and $D_{pV} = 1000$ nm. By using these values, the total uncertainty as a function of $kA$ ($A$ is the NW cross-sectional area) is calculated and shown in Fig. 4(b). For highly conductive NWs ($kA > 5 \times 10^{-14}$ WmK$^{-1}$), the measurement uncertainty is around 60%, indicating that obtaining an estimate of the thermal properties of a single NW is possible. However, this also indicates that it is *not* possible to measure an individual single-wall CNT with the supported platform because its $kA$ is low ($< 2 \times 10^{-14}$ WmK$^{-1}$), although it may be possible to measure one multi-wall CNT as long as the $kA$ condition above is satisfied.

Before concluding, we note that the supported thermometry platform is not limited to SiO$_2$/Si substrates, but could be extended to thermally insulating plastic substrates like Kapton, polyethylene terephthalate (PET), and so on. Here, we consider measurements with a 25 μm thick Kapton substrate on a heat sink[36] [inset of Fig. 5(a)]. We note that results for other Kapton thickness or PET are similar. Since the plastic substrate is generally tens of microns thick, its background thermal conductance will be typically larger than that of the sample; thus, the sample should be trimmed (etched), leaving just the portion between the heater and sensor [inset of Fig. 5(b)]; otherwise, it will be difficult to sense the difference between the sample measurement and the control experiment without the sample, resulting in a large uncertainty. By using $k_{ps} = 0.37$ Wm$^{-1}$K$^{-1}$ for Kapton (DuPont™ Kapton® MT) and $G_{\square} = 100$ nWK$^{-1}$ for the sample, we calculate the measurement uncertainty as a function of the heater-sensor distance $L_{HS}$, as shown in Fig. 5(a). The minimized uncertainty is ~ 38% at $L_{HS} = 1.2$ μm. With this optimized structure, we further calculate the uncertainty change as a function of the sample thermal sheet conductance [Fig. 5(b)], and find the smallest sensible $G_{\square}$ within a 50% error for a Kapton substrate is ~60 nWK$^{-1}$, which is ~2.5 times higher than for the optimized SiO$_2$/Si substrate. Correspondingly, the required density for CNT and NW arrays will be also 2.5 times higher, which remains achievable in experiments.

In conclusion, we demonstrated that a relatively simple, substrate-supported platform can be used to measure heat flow in nanoscale samples like graphene and CNT or NW arrays. This platform requires fewer fabrication efforts and is useful for materials that are difficult to suspend, but the sample thermal conductivity must be extracted by 3D finite element analysis. Based on careful uncertainty analysis, we find the platform design can be optimized and the smallest thermal sheet conductance measurable by this method within 50% error is estimated to be ~25 nWK$^{-1}$ at room temperature. This thermometry platform can also be applied to individual nanowires and can be implemented both on SiO$_2$/Si (or similar) and flexible plastic substrates.

We thank D. Estrada, B. Howe, and A. Bezryadin for experimental assistance. The work was supported by the Presidential Early Career (PECASE) award from the Army Research Office, the National Science Foundation (NSF), the Nanotechnology Research Initiative (NRI), and National Research Foundation of Korea (NRF-2012M3C1A1048861).

**REFERENCES**

1. D. G. Cahill, P. V. Braun, G. Chen, D. R. Clarke, S. Fan, K. E. Goodson, P. Keblinski, W. P. King, G. D. Mahan, A. Majumdar, H. J. Maris, S. R. Phillpot, E. Pop and L. Shi, Appl. Phys. Rev. **1** (1), 011305 (2014).

2. N. B. Li, J. Ren, L. Wang, G. Zhang, P. Hanggi and B. W. Li, Rev. Mod. Phys. **84** (3), 1045 (2012).

3. T. F. Luo and G. Chen, Phys. Chem. Chem. Phys. **15** (10), 3389 (2013).

4. E. Pop, Nano Res. **3** (3), 147 (2010).

5. D. G. Cahill, K. Goodson and A. Majumdar, J. Heat Transf. **124** (2), 223 (2002).

6. L. Shi, D. Y. Li, C. H. Yu, W. Y. Jang, D. Kim, Z. Yao, P. Kim and A. Majumdar, J. Heat Transf. **125** (5), 881 (2003).

7. F. Volklein, H. Reith, T. W. Cornelius, M. Rauber and R. Neumann, Nanotechnology **20** (32), 325706 (2009).

8. M. C. Wingert, Z. C. Y. Chen, S. Kwon, J. Xiang and R. K. Chen, Rev. Sci. Instrum. **83** (2), 024901 (2012).

9. R. Sultan, A. D. Avery, G. Stiehl and B. L. Zink, J. Appl. Phys. **105** (4), 043501 (2009).

10. J. K. Yu, S. Mitrovic, D. Tham, J. Varghese and J. R. Heath, Nat. Nanotechnol. **5** (10), 718 (2010).

11. P. Ferrando-Villalba, A. F. Lopeandia, A. Ll, J. Llobet, M. Molina-Ruiz, G. Garcia, M. Gerbolès, F. X. Alvarez, A. R. Goñi, F. J. Muñoz-Pascual and J. Rodríguez-Viejo, Nanotechnology **25** (18), 185402 (2014).

12. C. W. Chang, A. M. Fennimore, A. Afanasiev, D. Okawa, T. Ikuno, H. Garcia, D. Y. Li, A. Majumdar and A. Zettl, Phys. Rev. Lett. **97** (8), 085901 (2006).

13. J. K. Yang, Y. Yang, S. W. Waltermire, T. Gutu, A. A. Zinn, T. T. Xu, Y. F. Chen and D. Y. Li, Small **7** (16), 2334 (2011).

14. I. Jo, M. T. Pettes, J. Kim, K. Watanabe, T. Taniguchi, Z. Yao and L. Shi, Nano Lett. **13** (2), 550 (2013).

15. Z. Q. Wang, R. G. Xie, C. T. Bui, D. Liu, X. X. Ni, B. W. Li and J. T. L. Thong, Nano Lett. **11** (1), 113 (2011).

16. M. M. Sadeghi, I. Jo and L. Shi, Proc. Natl. Acad. Sci. USA **110** (41), 16321 (2013).

17. A. I. Boukai, Y. Bunimovich, J. Tahir-Kheli, J. K. Yu, W. A. Goddard and J. R. Heath, Nature **451** (7175), 168 (2008).

18. N. Stojanovic, J. M. Berg, D. H. S. Maithripala and M. Holtz, Appl. Phys. Lett. **95** (9), 091905 (2009).

19. W. Y. Jang, Z. Chen, W. Z. Bao, C. N. Lau and C. Dames, Nano Lett. **10** (10), 3909 (2010).

20. M.-H. Bae, Z. Li, Z. Aksamija, P. N. Martin, F. Xiong, Z.-Y. Ong, I. Knezevic and E. Pop, Nat. Commun. **4**, 1734 (2013).

21. M. M. Shulaker, J. Van Rethy, T. F. Wu, L. S. Liyanage, H. Wei, Z. Y. Li, E. Pop, G. Gielen, H. S. P. Wong and S. Mitra, ACS Nano **8** (4), 3434 (2014).

22. S. Islam, Z. Li, V. E. Dorgan, M.-H. Bae and E. Pop, IEEE Electron Device Lett. **34** (2), 166 (2013).

23. A. A. Balandin, Nat. Mater. **10** (8), 569 (2011).

24. E. Pop, V. Varshney and A. K. Roy, MRS Bulletin **37** (12), 1273 (2012).

25. Y. Xu, Z. Li and W. Duan, Small **10** (11), 2182 (2014).

26. See supplementary material at http://dx.doi.org/10.1063/1.4887365 for complimentary simulation results and details of uncertainty analysis.

27. The non-etched case has the similar results and is shown in the supplementary material (Ref. 26).

28. K. Kurabayashi, M. Asheghi, M. Touzelbaev and K. E. Goodson, J. Microelectromech. Syst. **8** (2), 180 (1999).

29. M. E. Itkis, F. Borondics, A. P. Yu and R. C. Haddon, Nano Lett. **7** (4), 900 (2007).

30. D. Estrada and E. Pop, Appl. Phys. Lett. **98** (7), 073102 (2011).

31. S. W. Hong, T. Banks and J. A. Rogers, Adv. Mater. **22** (16), 1826 (2010).

32. C. A. Wang, K. M. Ryu, L. G. De Arco, A. Badmaev, J. L. Zhang, X. Lin, Y. C. Che and C. W. Zhou, Nano Res. **3** (12), 831 (2010).

33. D. Y. Li, Y. Y. Wu, P. Kim, L. Shi, P. D. Yang and A. Majumdar, Appl. Phys. Lett. **83** (14), 2934 (2003).

34. A. I. Hochbaum, R. Chen, R. D. Delgado, W. Liang, E. C. Garnett, M. Najarian, A. Majumdar and P. Yang, Nature **451** (7175), 163 (2008).

35. M. C. McAlpine, H. Ahmad, D. W. Wang and J. R. Heath, Nat. Mater. **6** (5), 379 (2007).

36. In simulations, the ambient (isothermal) boundary condition must be applied to at least one surface of the substrate, but due to its low *k*, the outer surfaces of the plastic substrate cannot reach the ambient temperature. Thus, a heat sink with high *k* is added for this purpose. In practice, measurements are also generally performed in vacuum and on a temperature controlled stage, which serveses as a heat sink.

**FIGURES**

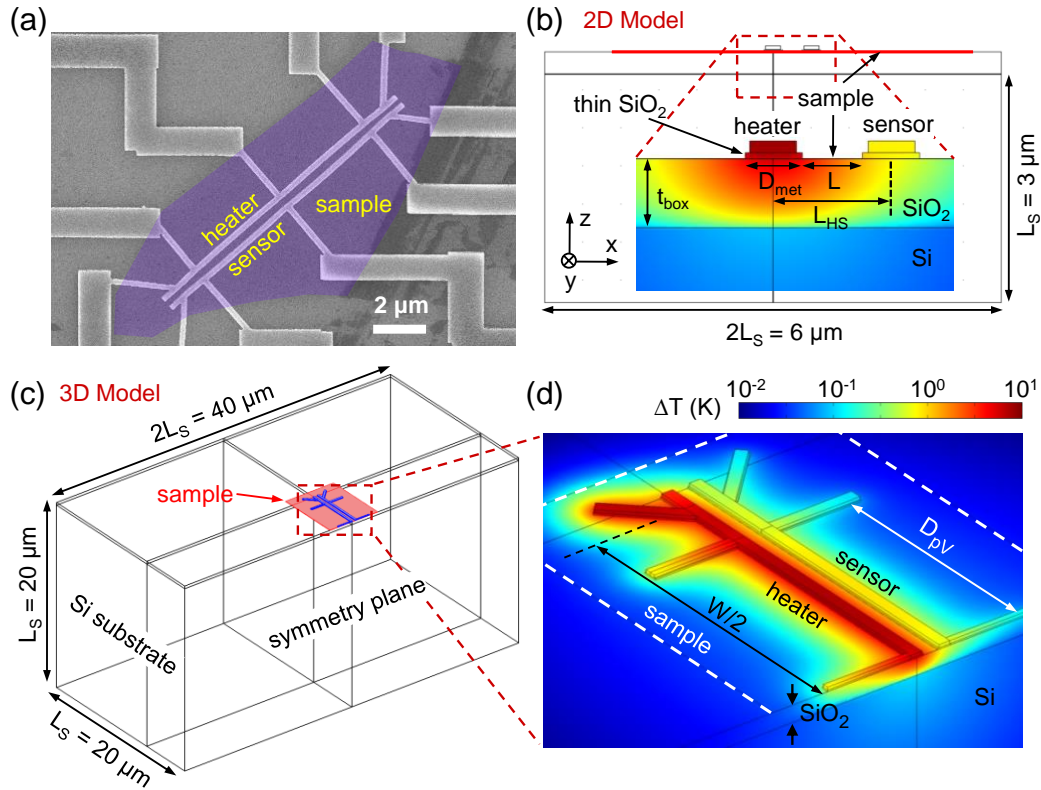

FIG. 1. (a) Scanning electron microscopy image of supported thermometry platform designed to measure thermal conductivity of a graphene sample (purple) on a $SiO_2$/Si substrate. (b) and (c) 2D and 3D finite element models used to simulate heat conduction in the supported thermometry platform, respectively. In the 2D model, only the cross-section is included and the zoom-in shows the typical temperature distribution with heating current applied through the heater. (d) Zoomed-in temperature distribution around heater and sensor obtained from 3D simulation, which matches with measured temperature. White dashed lines indicate the outline of the sample, which is highlighted by the red line and pink rectangle in (b) and (c), respectively. The detailed shape and size of the sample will not affect the simulation.

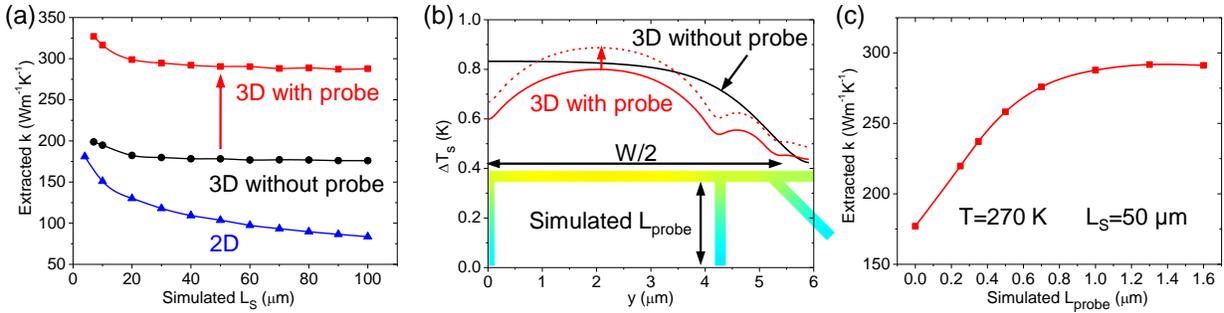

FIG. 2. (a) Extracted graphene thermal conductivity as a function of the Si substrate size $L_S$ for different models. (b) Simulated temperature profiles along the sensor from 3D models with and without probes. Two solid lines are obtained by using the same graphene $k$. Changing from solid to dash red lines corresponds to the red arrow in (a). (c) Extracted graphene $k$ converges as the simulated probe length $L_{probe}$ increases [corresponding to the red arrow in (a)].

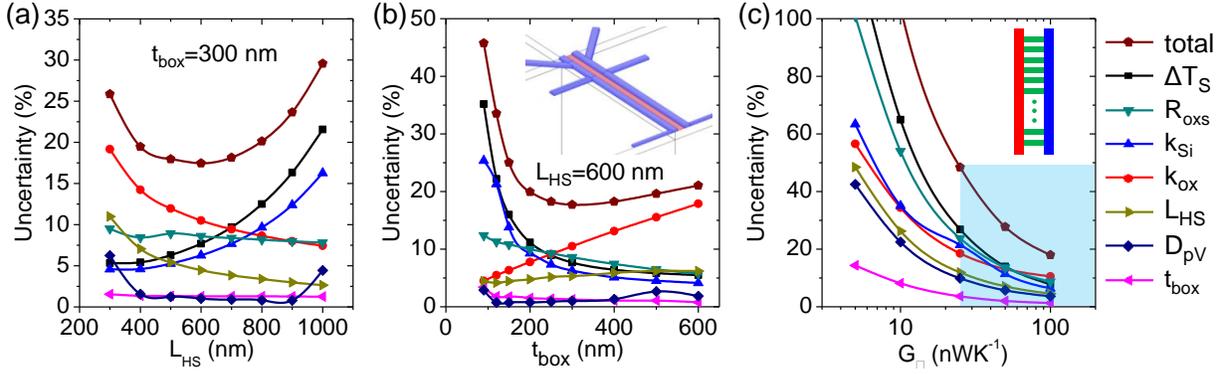

FIG. 3. (a) and (b) Estimated uncertainty of extracted sample thermal conductivity as a function of the heater-sensor midpoint separation $L_{HS}$ and the bottom oxide thickness $t_{box}$, respectively. This gives the optimized design with $L_{HS}$ = 600 nm and $t_{box}$ = 300 nm for the partly etched sample case [inset of (b), pink region indicates the sample]. (c) Estimated uncertainty of extracted sample $k$ increases as its thermal sheet conductance $G_\square$ decreases for $L_{HS}$ = 600 nm and $t_{box}$ = 300 nm, showing the measureable $G_\square$ (blue region) by this SiO$_2$/Si supported thermometry platform. Inset is a schematic for CNT/NW array measurements, where green lines are CNTs/NWs, and red and blue lines are heater and sensor, respectively.

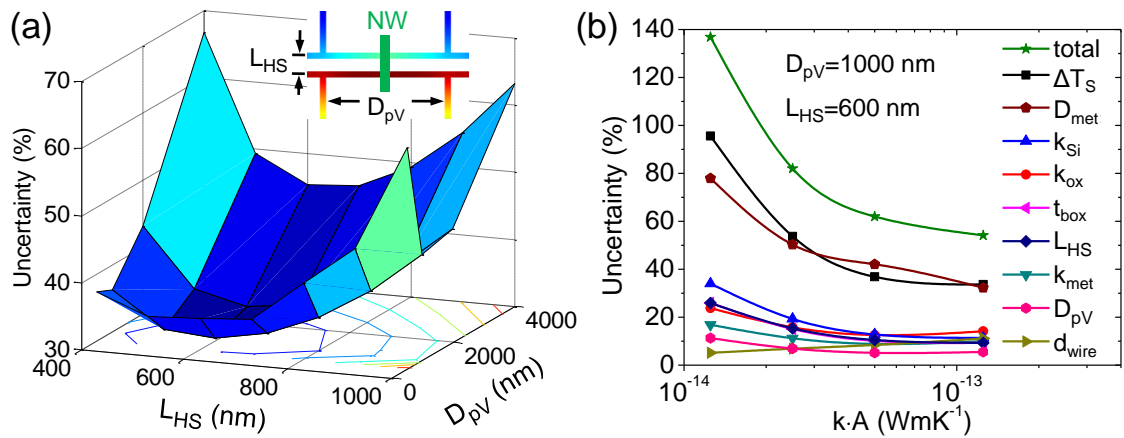

FIG. 4. (a) Optimizing the design for measuring a single nanowire (see inset) by estimating the uncertainty of extracted nanowire thermal conductivity as a function of $L_{HS}$ and $D_{pV}$. Here, only the uncertainty contributed from the measured sensor temperature rise ($\Delta T_S$) is calculated. The optimized design is $L_{HS}$ = 600 nm and $D_{pV}$ = 1000 nm. (b) Estimated uncertainty of extracted nanowire $k$ increases as its cross-sectional thermal conductance $kA$ decreases.

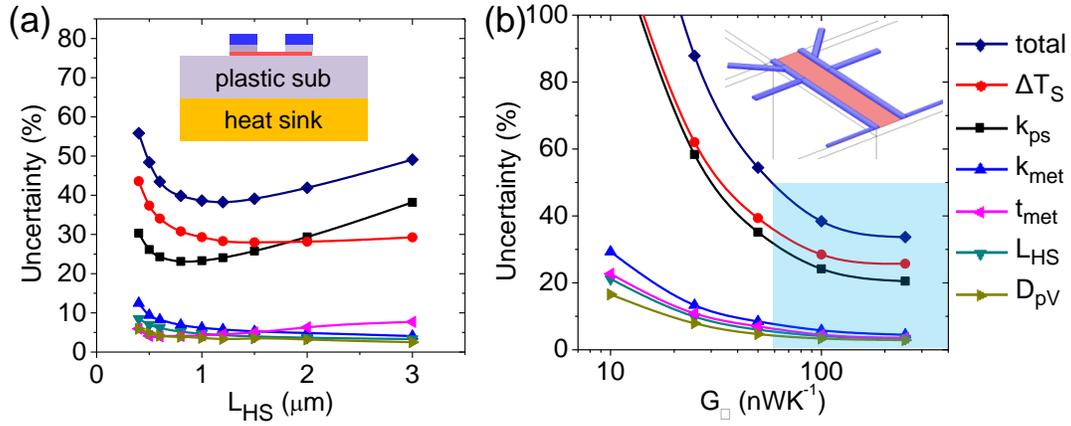

FIG. 5. (a) Optimizing the heater-sensor midpoint separation $L_{HS}$ with the platform supported by a plastic substrate on a heat sink[36] (see inset). The estimated uncertainty of extracted sample $k$ is minimized at $L_{HS} = 1.2$ μm for a 25 μm thick Kapton substrate. (b) Estimated uncertainty of extracted sample $k$ increases as its thermal sheet conductance $G_{\square}$ decreases, giving the sensible range (blue) of $G_{\square}$ by this platform. Inset shows the optimized structure of the platform applied to the plastic substrate, with $L_{HS} = 1.2$ μm and the sample (red) only between heater and sensor.

## Supplementary Materials for

# Substrate-supported thermometry platform for nanomaterials like graphene, nanotubes, and nanowires

Zuanyi Li[1,2], Myung-Ho Bae[3], and Eric Pop[1*]

[1]Electrical Engineering, Stanford University, Stanford, CA 94305, USA
[2]Dept. of Physics, Univ. Illinois, Urbana-Champaign, IL 61801, USA
[3]Korea Research Institute of Standards and Science, Daejeon, 305-340, Republic of Korea
[*]Contact: epop@stanford.edu

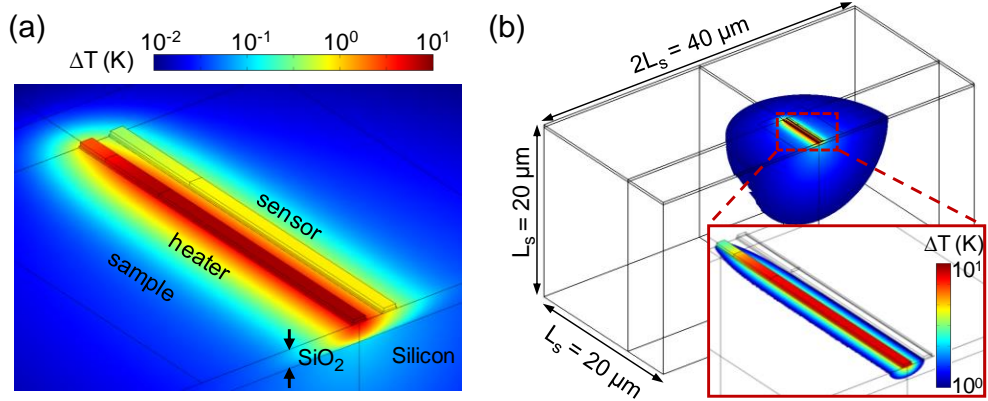

FIG. S1. (a) Structure and temperature distribution for the 3D model without probe arms, which can be regarded as the one extruded from the 2D model, and hence has a better correspondence to the 2D model than the 3D with probe model. (b) Evolvement of temperature isosurface shape in 3D simulation. When heat spreads out from the heater, the isosurface is close to a cylinder at the beginning (inset), behaving as 2D heat conduction, while it changes to a sphere after ~10 μm from the center, indicating heat spreading along the third direction cannot be ignored, which is the reason why the 2D model fails.

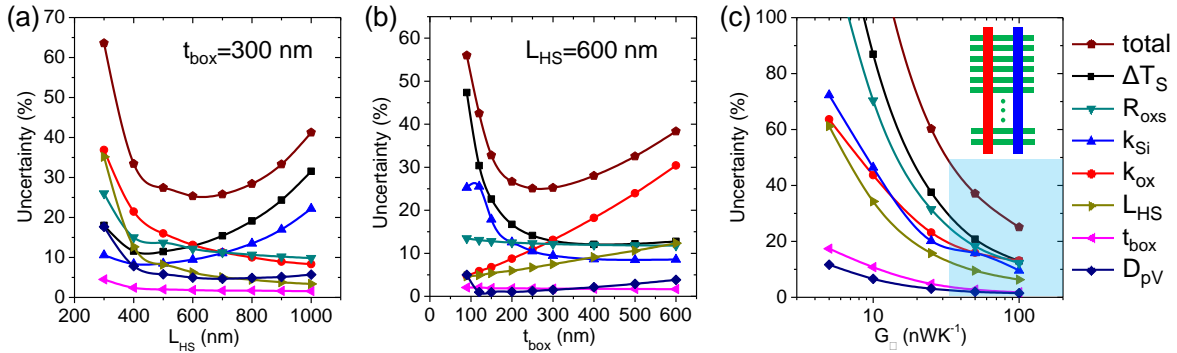

FIG. S2. Results for the sample non-etched case (see Fig. 1). (a),(b) Optimize the heater-sensor distance $L_{HS}$ and the bottom oxide thickness $t_{box}$, respectively, by estimating uncertainty change in the sample thermal conductivity extraction. The optimized $L_{HS}$ and $t_{box}$ are almost the same as those for the sample non-etched case [inset of Fig. 3(b)], but with a slightly higher uncertainty ~25%. (c) Estimated uncertainty of extracted sample $k$ increases as its thermal sheet conductance $G_\square$ decreases for $L_{HS}$ = 600 nm and $t_{box}$ = 300 nm, showing the measureable $G_\square$ within a 50% error (blue region) is $G_\square \geq 32$ nWK$^{-1}$. Inset is a schematic for CNT/NW array measurements, where green lines are CNTs/NWs, and red and blue lines are heater and senor, respectively.

| Input parameters (T = 270 K) | | | Units | Values $x_i$ | Uncertainty $u_{xi}$ | $u_{xi}/x_i$ | Sensitivity $s_i$ | Contribution $c_i=|s_i|\times u_{xi}/x_i$ | $c_i^2/\Sigma c_i^2$ |
|---|---|---|---|---|---|---|---|---|---|
| Expt. | Sensor response | $\Delta T_S/P_H$ | K/µW | 0.01623 | 0.0003 | 1.8% | 4.41 | 8.2% | 15.4% |
| | Heater response | $\Delta T_H/P_H$ | | 0.14080 | 0.001 | 0.7% | 2.42 | 1.7% | 0.7% |
| Thermal | Thermal conductivity of SiO$_2$, Si, metal | $k_{ox}$ | W/m/K | 1.213 | 0.04 | 3.3% | 4.32 | 14.2% | 46.9% |
| | | $k_{Si}$ | | 127 | 15 | 11.8% | 0.46 | 5.5% | 6.9% |
| | | $k_{met}$ | | 49 | 4 | 8.2% | 0.40 | 3.3% | 2.5% |
| | TBR of graphene/SiO$_2$, SiO$_2$/Si, metal/SiO$_2$ interfaces | $R_{gox}$ | m$^2$K/W | 1.15E-08 | 2.0E-09 | 17.4% | -0.11 | 1.8% | 0.8% |
| | | $R_{oxs}$ | | 9.99E-09 | 3.5E-09 | 35.0% | -0.25 | 8.8% | 17.7% |
| | | $R_{mox}$ | | 1.02E-08 | 3.5E-09 | 34.3% | 0.06 | 2.0% | 0.9% |
| Geometrical | Thickness of bottom and top SiO$_2$, and metal | $t_{box}$ | nm | 294 | 1 | 0.3% | -5.90 | 2.0% | 0.9% |
| | | $t_{tox}$ | | 25 | 1 | 4.0% | -0.02 | 0.1% | 0.0% |
| | | $t_{met}$ | | 50 | 2 | 4.0% | 0.40 | 1.6% | 0.6% |
| | Distance of H/S | $L_{HS}$ | | 505 | 5 | 1.0% | 5.09 | 5.0% | 5.9% |
| | Width of metal and top SiO$_2$ lines | $D_{met}$ | | 199 | 4 | 2.0% | -0.15 | 0.3% | 0.0% |
| | | $D_{tox}$ | | 243 | 4 | 1.6% | -0.31 | 0.5% | 0.1% |
| | Half length of H/S metal lines | $L_{met}/2$ | µm | 5.9 | 0.04 | 0.7% | 0.19 | 0.1% | 0.0% |
| | Distance of 2 Voltage probes | $D_{pV}$ | | 4.21 | 0.02 | 0.5% | 3.58 | 1.7% | 0.7% |
| | Distance of Current and Voltage probes | $D_{pIV}$ | | 1.04 | 0.02 | 1.9% | -0.12 | 0.2% | 0.0% |

Table S1. Calculated sensitivities and uncertainty analysis for the extracted graphene $k$ from our measurement at 270 K [correspond to the converged value in Fig. 2(c) with $L_S$ = 50 µm and $L_{probe}$ = 1.6 µm]. The total uncertainty is ~20.8%, and it mainly arises from the contributions ($c_i \geq 5\%$) of the measured sensor response ($\Delta T_S$), thermal conductivity of bottom SiO$_2$ ($k_{ox}$) and Si substrate ($k_{Si}$), thermal boundary resistance (TBR) of SiO$_2$/Si interface ($R_{oxs}$), and heater-sensor distance ($L_{HS}$). The values and their uncertainties of $k_{ox}$ and $R_{oxs}$ are from our previous measurements.[1] The thermal conductivity of metal lines ($k_{met}$) is calculated from the measured electrical resistance according to the Wiedemann-Franz Law. $k_{Si}$ is well-known data from Ref. S2. The TBR values and their uncertainties for the graphene/SiO$_2$ interface and metal/SiO$_2$ interface are based on measurements in Refs. S3 and S4, respectively. These TBRs have small contributions to the total uncertainty ($c_i \leq 2\%$) due to their small sensitivities ($s_i$). Even if their values change by a factor of 2 (i.e., $u_{xi}$=100%), their contributions become $c_i$ = 10% and 6%, which only change the total uncertainty from ~20.8% to ~23% and ~21.5%, respectively.

**References**

S1. Bae, M.-H. et al. Ballistic to diffusive crossover of heat flow in graphene ribbons. *Nat. Commun.* **4**, 1734 (2013).

S2. McConnell, A.D. & Goodson, K.E. Thermal conduction in silicon micro- and nanostructures. *Annual Reviews of Heat Transfer* **14**, 129 (2005).

S3. Chen, Z., Jang, W., Bao, W., Lau, C.N. & Dames, C. Thermal contact resistance between graphene and silicon dioxide. *Appl. Phys. Lett.* **95**, 161910 (2009).

S4. Koh, Y.K., Bae, M.-H., Cahill, D.G. & Pop, E. Heat Conduction across Monolayer and Few-Layer Graphenes. *Nano Lett.* **10**, 4363 (2010).